\documentclass[aps,epsfig,prl,floats,superscriptaddress,twocolumn,showpacs]{revtex4}%
\usepackage{epsfig}
\usepackage{dcolumn}
\usepackage{bm}
\usepackage{amsmath}
\usepackage{amsfonts}
\usepackage{amssymb}
\usepackage[latin1]{inputenc}%
\newcommand{\vQ}{\textbf{Q}}
\renewcommand{\vr}{\textbf{r}}
\newcommand{\vk}{\textbf{k}}
\newcommand{\vR}{\textbf{R}}
\renewcommand{\Re}{\textrm{Re}}
\renewcommand{\Im}{\textrm{Im}}
\begin{document}
\title{Arrested  Kondo effect and hidden order in URu$_2$Si$_2$}
\author{Kristjan Haule and Gabriel Kotliar }

\affiliation{Center for Materials Theory, Serin Physics
Laboratory, Rutgers University, 136 Frelinghuysen Road,
Piscataway, New Jersey 08854, USA.}

\pacs{71.27.+a, 72.15.Rn, 71.30.+h}

% Intro of 200-300 words
% 1500 words of the rest of the text
% (References, title, author list and acknowledgements do not have to
%   be included in total word counts.)
% 4pages - 1300 words per page

% <= 30 references
% Methods: in brief (less than 200 words) can be in the
% text. Otherwise Methods-Summary  (of 300 words) and Methods of 1000
% words.

% Science
% Reports (up to ~2500 words or ~3 journal pages) present important new
% research results of broad significance. Reports should include an
% abstract, an introductory paragraph, up to four figures or tables, and
% a maximum of 30 references. Materials and Methods should usually be
% included in supporting online material, which should also include
% information needed to support the paper's conclusions.

%%%% Now: 2300 of all words

\maketitle

%\begin{abstract}
% \textbf{
% Complex electronic matter exhibit subtle forms of self organization
% whose detection pose enormous challenges to the physicists.  For lack
% of a better name, form of order that are known to exist but so far
% have resisted identification are called "hidden orders".
% %
% %  
% One prominent example is provided by the actinide based heavy fermion
% material URu$_2$Si$_2$.
% %
% %
% Here we develop a first principles theoretical method
% %
% to analyze the electronic spectrum of correlated materials as a
% function of the position inside the unit cell of the crystal, and use
% it to identify the low energy excitations of the URu$_2$Si$_2$.
% %
% We identify the order parameter of the hidden order state, and show
% that it is intimately connected with magnetism. Our findings explain
% why is the "hidden order" so hard to detect experimentally.
% %
% We show the temperature evolution of the electronic states of the
% material. At temperature below 70$\,$K U-$5f$ electrons undergo a
% multichannel Kondo effect, which is arrested at low temperature by the
% crystal field splitting.
% %
% At even lower temperature, we stabilized two broken symmetry states
% characterized by a complex order parameter $\psi$. A real
% $\psi$ describes the hidden order phase, and an imaginary $\psi$
% corresponds to the large moment antiferromagnetic phase, thus
% providing a unified picture of the broken symmetry phases, which are
% realized in this material.
% }
% %\end{abstract}
%\maketitle

\textbf{
Complex electronic matter exhibit subtle forms of self organization
which are almost invisible to the available experimental tools, but
which have dramatic physical consequences.
One prominent example is provided by the actinide based heavy fermion
material URu$_2$Si$_2$. At high temperature, the U-$5f$ electrons in
URu$_2$Si$_2$ carry a very large entropy. This entropy is released at
$17.5\,$K via a second order phase transition \cite{first} to a state
which remains shrouded in mystery, and which was termed a "hidden
order" state~\cite{Coleman}.
Here we develop a first principles theoretical method
to analyze the electronic spectrum of correlated materials as a
function of the position inside the unit cell of the crystal, and use
it to identify the low energy excitations of the URu$_2$Si$_2$.
We identify the order parameter of the hidden order state, and show
that it is intimately connected with magnetism.
%Our findings explain
%why is the "hidden order" so hard to detect experimentally.
%
We present first principles results for the temperature evolution of
the electronic states of the material. At temperature below 70$\,$K
U-$5f$ electrons undergo a multichannel Kondo effect, which is
arrested at low temperature by the crystal field splitting.
At lower temperatures, two broken symmetry states emerge, characterized
by a complex order parameter $\psi$. A real $\psi$ describes the
hidden order phase, and an imaginary $\psi$ corresponds to the large
moment antiferromagnetic phase, thus providing a unified picture of
the two broken symmetry phases, which are realized in this material.
}

%\section{\bf Intro on  heavy fermions and URu2Si2.}

%We present ab-initio Dynamical Mean Field Theory calculations for this
%material and we show how the U-$5f$ electrons undergo a multichannel
%Kondo effect, which is arrested at low temperature by the crystal
%field splitting.

\begin{figure}[t]
\begin{center}
\includegraphics[width=0.29\linewidth]{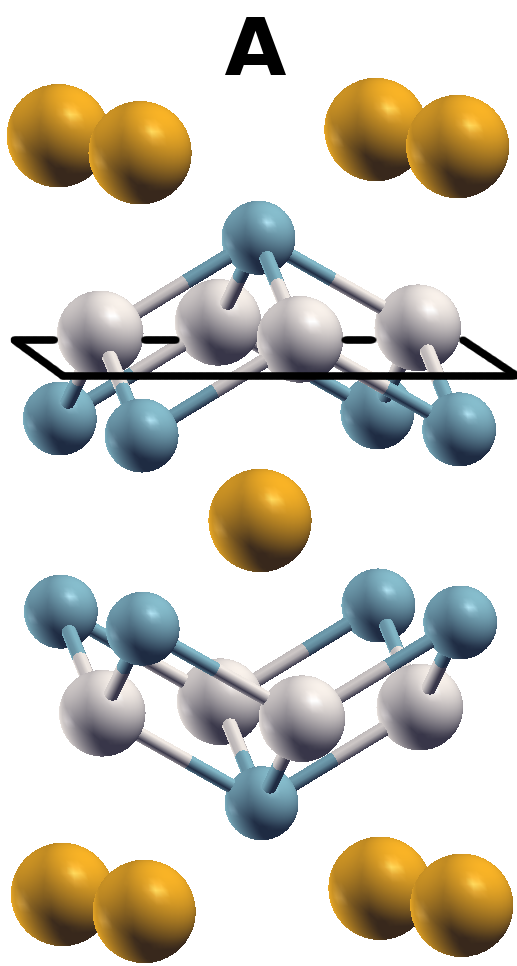}
\includegraphics[width=0.69\linewidth]{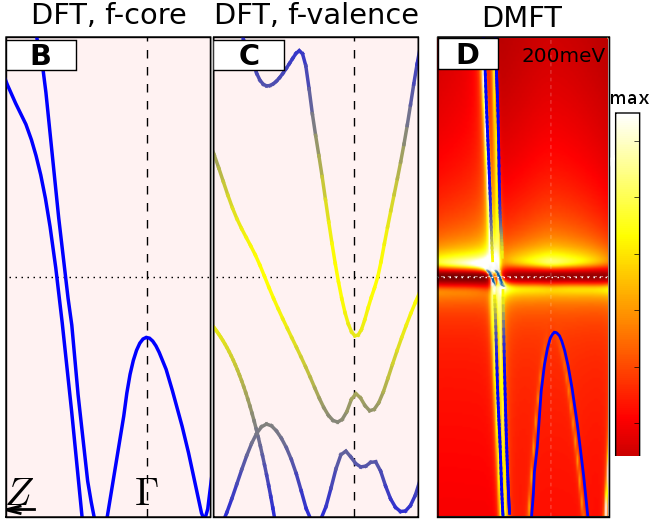}
\end{center}
\caption{ \textbf{(A)} Crystal structure of URu$_2$Si$_2$. Gold, grey
  and green spheres correspond to U, Ru, and Si, respectively. The
  black box marks the cut of the unit cell for which we show
  electronic states in Fig.~\ref{fig2}.  \textbf{(B,C,D)} A sketch of
  the arrested Kondo effect phenomena. In the vertical axis we display
  the energy (in the interval [$-200\,$meV, $200\,$meV]) and in the
  horizontal axis the momentum (around the center of the Brillouin
  zone). The horizontal dotted line marks the Fermi level.  The light
  $spd$ bands are plotted by blue lines, while the bright colors
  correspond to the primarily U-$f$ spectra.  In standard band
  structure calculation, the heavy $U$ bands are either excluded from
  the calculation (shown in \textbf{B}), or included in the one
  electron theory (\textbf{C}).  The many body calculation
  (\textbf{D}) shows the new phenomena of Kondo effect developing only
  at intermediate energy scales and intermediate temperatures $\Delta
  \lesssim T \lesssim T^*$, where $\Delta\sim 35K$ is the crystal
  field splitting and $T^*\sim 70 K$ is the coherence scale. The
  temperature in panel (\textbf{D}) is $T=19\,$K$<\Delta$ and
  color coding of the spectra is the projection of the total DMFT
  electronic spectra to the U-$5f$ character.  The DMFT method
  predicts the U $5f$ occupation to be 2.0, hence U-$5f$ configuration
  in solid is $U^{4+}$.  }
\label{fig1}
\end{figure}

%
%In spite of decades long efforts (for a recent review see
%Ref.~\cite{Coleman})
%the basic issues surrounding the electronic structure of this
%compound, such as the valence of $U$, or the degree of localization or
%itinerancy of the $U-5f$ electrons have not been understood so far,
%preventing the resolution of the hidden order conundrum.
%%%% 103 words

% Here we study  the electronic structure of this material using 
% first principles methods to treat the strong correlations.
% %
% We develop a new theoretical tool to analyze the electronic spectrum
% as a function of the position inside the unit cell of the crystal, and
% use it to identify the low energy excitations of the system as
% fluctuations between the singlet atomic ground state of the $U-5f^2$
% configuration and the singlet first excited state of the same
% configuration, separated from each other by only $35\,$K. We identify
% the order parameter of the hidden order state, and show that it is
% intimately connected with magnetism.  This picture explains why is the
% "hidden order" state rare in nature, and why it is so hard to detect
% its order parameter.

URu$_2$Si$_2$ crystallizes in the body centered tetragonal structure
displayed in Fig.~\ref{fig1}A.  Starting form a localized point of
view for $U$-$5f$ electrons, i.e., treating them as core states in
density functional theory calculations, one obtains a set of wide
bands, entirely of $spd$ character, as shown in Fig.~\ref{fig1}B.
In an itinerant picture, there are $f$ states close to the Fermi level
which hybridize with the $spd$ bands to form itinerant states of
predominantly $f$ character, with a bandwidth of the order of
$1\,$eV. This is shown in Fig.~\ref{fig1}C.  This situation is
realized in Ce and Yr based heavy fermion materials, whose electronic
states are well described by narrowing the bands of the density
functional theory by a factor of 10-1000, to account for the heavy
mass \cite{Allen}.

For URu$_2$Si$_2$, we propose a new scenario for the transfer of
atomic $f$ weight to the itinerant carriers which we dub the "arrested
Kondo effect". At high temperatures, above the characteristic
coherence temperature $T^* \sim 70\,$K, the U-$5f$ electrons are
localized, and do not participate in forming the low energy bands.
The $U$ atoms settle in $5f^2$ configuration, for which the crystal
environment chooses the non-degenerate atomic ground state.
However, the first excited state, which is also non-degenerate, is
only $\Delta=35\,$K above the ground state, as first observed in
polarized neutron scattering experiments \cite{Broholm}.  Hence in the
temperature range $\Delta<T<T^*$, the ground state seems doubly
degenerate, hence the Kondo effect develops, leading to formation of
very narrow states near the Fermi energy, and a narrow peak in the
density of states. The Kondo effect is partially arrested below
$T<\Delta$, because the crystal fields splitting between the two
singlet states induces a partial gaping at the Fermi level. The
situation is shown in Fig.~\ref{fig1}C. A non-dispersing slab of
$f$ spectral weight is pushed to roughly $8\,$ meV away from the Fermi
energy. While only a small fraction of the $f$ spectral weight is
present at the Fermi level, it has a significant effect, resulting in
a mass enhancement factor which is over 200 at $T=19\,$K.  For
temperatures above the crystal field splitting energy, $T\sim \Delta$,
the electronic scattering rate is anomalously large.
This is a signature of multichannel Kondo physics, whereby the two
singlets play the role of the spin in the Kondo problem that scatters
degenerate bands. The relevance of the two channel Kondo physics to
$U$ based impurity models was first proposed by D. Cox
\cite{TwoChannel}. Indeed experimental data on dilute
U$_x$Th$_{1-x}$Ru$_2$Si$_2$ systems are well described by the two
channel Kondo model \cite{amitsuka}. It is remarkable that the Fermi
liquid regime in URu$_2$Si$_2$ can never be reached.
Once the two channel Kondo effect is arrested by the crystal field
splitting, long range order, driven by intersite effects, preempts the
formation of a Fermi liquid state with a low coherent scale.
% 
% the system transitions into an ordered state, because the
% intersite effects result in a higher energy gain than the formation of
% such a Fermi liquid state.
%%% number of words 473

\begin{figure}[t]
\begin{center}
\includegraphics[width=1.1\linewidth]{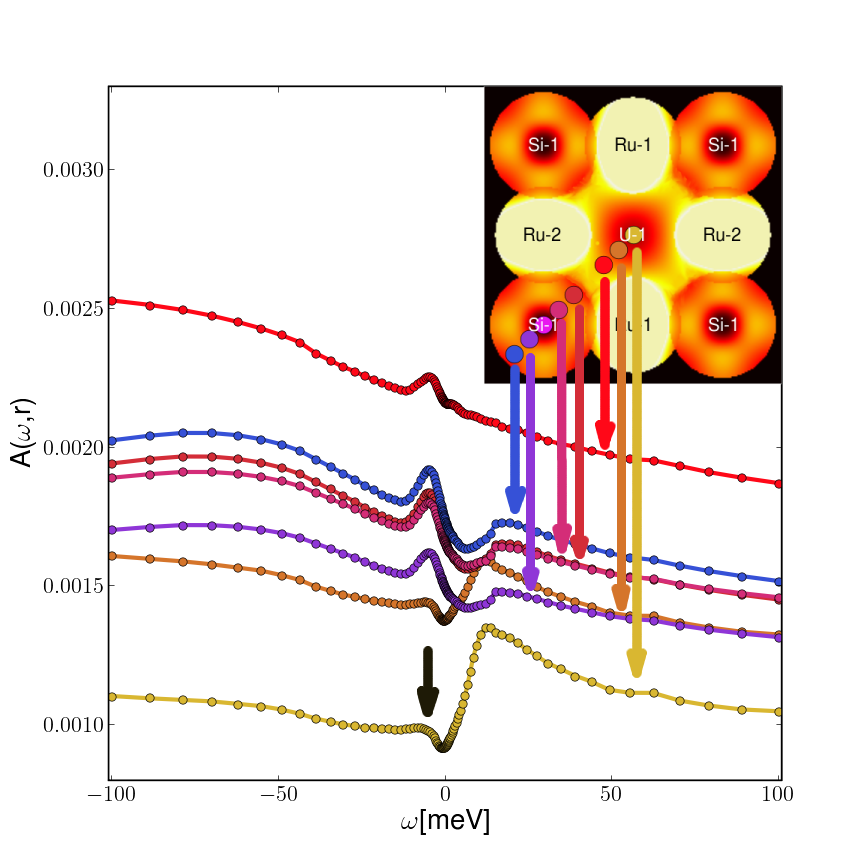}
\end{center}
\caption{ (inset) The calculated electronic density $A(\vr,\omega=0)$ in real
  space at $20\,$K, as measured by scanning tunneling microscopy
  experiments. We take a representative cut in real space, sketched in
  Fig.1, which cuts the $U$ muffin-tin sphere at $3/4$ height,
  guaranteeing a large matrix elements with the valence $4f$-electrons
  on $U$-atom. The main panel shows the energy dependence of the
  electronic density of states at various positions in real
  space. For each colored dot we show a curve in the main panel, which
  has the same color as the dot.}
\label{fig2}
\end{figure}
The one electron spectral function $A(\vr,\omega)$  represents 
the quantum mechanical probability for adding or removing an electron
with energy $\omega$ at a point  $\vr$ in real space.  We calculate it using
the combination of the density functional theory and dynamical mean
field theory (DMFT) \cite{RMP} 
\begin{equation}
  A(\vr,\omega) = \sum_\vk \chi^*_{\vk \alpha} (\vr)
  \left[G_{\vk,\alpha\beta}(\omega)-G_{\vk,\beta\alpha}^*(\omega)\right]\chi_{\vk \beta} (\vr)
  \label{green_function}
\end{equation}
where $\chi_\alpha(\vr)$ are the basis functions, and
$G_{\alpha\beta}$ is the electron Green's function expressed in the basis of
$\chi$.  More details of the implementation of the DMFT method for this
problem is given in the online material.% and Ref.~\onlinecite{LongPaper}. 

These computational studies are the theoretical counterpart of the
scanning tunneling spectroscopy technique, that has been very useful
in describing the properties of numerous correlated materials
\cite{Davis} and was very recently applied to URu$_2$Si$_2$
\cite{Seamus}.
%%%  72 words

The evolution of the electron spectral function $A(\vr,\omega)$ along
the cut above $U$ atom in the unit cell, as depicted in
Fig.~\ref{fig1}A, is displayed in Fig.~\ref{fig2}.
The curves have an asymmetric lineshape of the type
$$A(\omega) \propto [(q^2-1) + 2 q (\omega/\Gamma)]/[(\omega/\Gamma)^2 + 1]$$
where $\Gamma$ is the width, and $q$ measures the asymmetry of the
lineshape. These curves were first introduced by Fano~\cite{fano}, to describe
scattering interference between a discrete state and a degenerate
continuum of states.
The continuum of states in scanning tunneling microscopy on
URu$_2$Si$_2$ is provided by itinerant $spd$ bands, and the narrow
discrete states are the U-$5f$ electronic states, depicted with
bright colors in Fig.~\ref{fig1}D.
Notice that the lineshape has the most characteristic Fano shape when
the position $\vr$ is on the $U$ or $Si$ atom, with positive asymmetry
$q>0$ on $U$ atom and negative asymmetry $q<0$ on $Si$ atom.  With the
exception of the small peak, marked with black arrow in
Fig.~\ref{fig2}, the lineshape on $U$ atom can be well fitted to the
Fano lineshape with parameter $q=1.24$ and $\Gamma=6.82$meV.

The main features of our calculations, including the characteristic
lineshape, the width, and the strength of the asymmetry of the Fano
lineshape, as well as the additional small peak around $6\,$meV, were
recently measured in scanning tunneling experiments by J.C. Seamus
Davis group \cite{Seamus}, the first experiment of this type on a heavy
fermion system.
%%% 236 words

\begin{figure}[t]
\begin{center}
\includegraphics[width=1.02\linewidth]{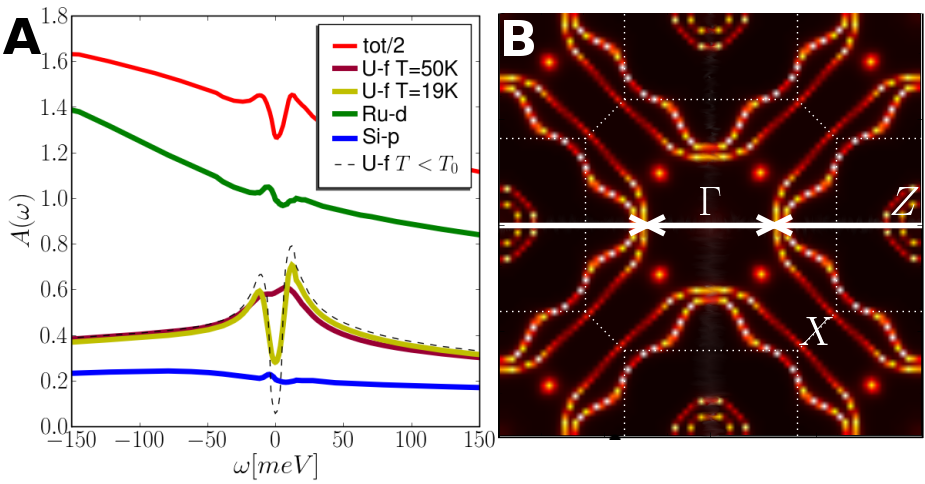}
\end{center}
\caption{ \textbf{(A)}
  The total density of electronic state (red) and
  the orbitally resolved components at $T=19K$. Also shown is the
  U-$f$ component at higher $T=50\,K$. Using Fermi liquid
  estimate for the specific hear coefficient
  $\gamma\propto A(\omega=0)/Z$ we obtain
  $\gamma\sim 85\,$ mJ/(mol K$^2$) at $T=50\,$K ($1/Z\sim 60$)
  and
  $\gamma\sim 140\,$mJ/(mol K$^2$) at $T=19\,$K ($1/Z\sim 200$).
\textbf{(B)} The DMFT Fermi surface of URu$_2$Si$_2$ in the paramagnetic state
  $T=20K >T_{order}$. We take a two dimensional horizontal cut in the the
  tetragonal body centered Brillouin zone passing through the
  origin.
}
\label{fig3}
\end{figure}
To understand the origin of the small peak at $6.8\,$meV, we integrate
the spectra over the whole unit cell and resolve it in the different
angular momentum channels of the different atoms.  This orbitally
resolved spectra at low energies is shown in Fig.~\ref{fig3}A.

While there are plenty of $Si$-$3p$ and $Ru$-$4d$ state at the Fermi level,
these states are only weakly energy dependent in $300\,$meV interval.
On the other hand, the heavy U-$5f$ electron states have a dip and a peak
at the same energy of $6.8\,$meV, which is clearly visible in the real
space spectra in Fig.~\ref{fig2}.
Therefore it is the U-$5f$ non-lorentzian shape of the electronic
density of state, which is responsible for the additional peak and the
asymmetry variation in the unit cell.
This dip-peak lineshape is also the key to unravelling the puzzle of
the electronic structure of the material, since it shows that instead
of a regular lorentzian Kondo resonance, we have a double peak at low
temperature, with a pseudogap at zero energy. This is the signature of
the Kondo effect, loosing its strength at temperature $T$ below the crystal
field splitting energy $T<\Delta$.
Below the ordering temperature (the ordering will be discussed below),
the pseudogap in the partial U-$5f$ spectra is considerably enhanced,
as presented in Fig.~\ref{fig3}A. Note that due to mixing of $spd$ and
$f$ states, there is finite admixture of U-$5f$ state at the Fermi
level, giving some "heaviness" to the quasiparticle even below the ordering
temperature.

The electronic states can also be resolved in momentum space by
computing the spectral function $A(\vk,\omega)$. The active state at
the Fermi level, given by peaks of $A(\vk,\omega=0)$, determine
the Fermi surface of the material, which is displayed in
Fig.~\ref{fig3}B.

For many Ce based heavy fermion compounds, the Fermi surface of
itinerant DFT calculation, as well as DMFT calculation, accounts
for the experimentally measured de-Hass van Alphen frequencies.
On the other hand, the DMFT Fermi surface of URu$_2$Si$_2$, shown in
Fig.~\ref{fig3}B, is qualitatively similar to the localized DFT
calculation, with $5f$'s excluded from the valence bands
\cite{Denlinger}.
%
%This can be understood as a consequence of the Luttinger's theorem,
%which states that the volume of the Fermi surface of the interacting
%and the non interacting electron system contains the same number of
%electrons modulo 2. For a Ce $f^1$ system, this implies that the
%itinerant DFT Fermi surface and the physical Fermi surface have the
%same volume, and this differs from the the localized DFT volume.
Note that the Luttinger's theorem, which counts the number of
electrons modulo 2, does not constraint the Fermi surface of the
material. Hence the Fermi surface does not need to resemble the
itinerant DFT result. 

The first principles DMFT calculations demonstrate that for
URu$_2$Si$_2$ the $5f^2$ configuration
%,
has the dominant weight. This has
important consequences, since it allows the physical Fermi surface to
have the same volume as the system with no $f$ electrons, such as
ThRu$_2$Si$_2$ \cite{DenlingerNew}.

Finally, the DMFT Fermi surface is hole-like, and the large hole Fermi
surface centered at $Z$ displays characteristic wave vectors
$(0.6,0,0)a^*$ and $(1.4,0,0)a^*$ shown in Fig.~\ref{fig3}B. Recent
neutron scattering experiments display low energy spectral weight at
these incomensurate wave vectors~\cite{Broholm2,Neutrons,Wiebe}.

%The electronic excitations with the wave vectors $(0.6,0,0)a^*$ and
%$(1.4,0,0)a^*$ were recently detected in neutron scattering
%experiments \cite{Broholm2,Neutrons,Wiebe}.

In the long range order phase, the Fermi surface substantially
reconstructs, and multiple small electron and hole pockets appear,
making this system a compensated metal, with the same number of hole
and electron carriers.

%%% 519 words

%\section{\bf Crystal Fields}

The $U$-$5f^2$ configuration has the total angular momentum $J=4$ and
is split into five singlets and two doublets in the tetragonal crystal
environment of URu$_2$Si$_2$.
The relative energy of these crystal field levels is still an open
problem and several sequences have been proposed. For a recent
review see Ref.~\cite{fazekas}.

The wave functions of the U-$5f^2$ configuration, with the largest
weight in the DMFT density matrix, are
\begin{eqnarray}
|\emptyset \rangle &=& \frac{i}{\sqrt{2}}(|4\rangle -|-4\rangle)\\
|1\rangle &=& \frac{\cos(\phi)}{\sqrt{2}}(|4\rangle +
|-4\rangle)-\sin(\phi)|0\rangle
\end{eqnarray}
Here $|J_z \rangle=|J=4,Jz\rangle$ is a two particle state of the
$J=4$ multiplet. We obtained $\phi\sim 0.23\pi$.  The separation
between the two singlets is of the order of $35\,$K. The probability
for the atomic ground state, $|\emptyset \rangle$, and the first excited state,
$|1\rangle$, at $20\,$K is 0.54 and 0.1, respectively.
%%% 117 words

The transition into ordered states requires an understanding of the
collective excitations, which are bound states of particle hole pairs.
The identification of the low lying singlets leads us to consider  the
following order parameter for  URu$_2$Si$_2$
\begin{equation}
\psi_i  =  \langle X_{\emptyset 1}(\vR_i) \rangle
\end{equation}
where $X_{\emptyset 1}=|\emptyset \rangle\langle 1|$ is the Hubbard operator which
measures the excitonic mixing between the two lowest lying $U$-$5f$
singlets at lattice site $\vR_i$.

This order parameter is complex. Its real part is proportional to the
hexadecapole operator of A$_{2g}$ irreducible representation of the
tetragonal symmetry,
$\Re\psi\propto \langle(J_x J_y + J_y J_x)(J_x^2-J_y^2 \rangle $,
introduced in the context of nuclear physics long
ago.
Its imaginary part is proportional to the magnetization along the $z$
axis, $\Im\psi\propto\langle J_z \rangle$, which is the only direction
allowed within this crystal field set of states. Experimentally, the
moment indeed points in $z$ direction \cite{Broholm}.

At low temperatures, we found two different stable DMFT solutions,
describing ordered states with non zero staggered $\langle\psi\rangle$
with wave vector $\vQ=(0,0,1)$.  The first solution has
$\langle\psi\rangle$ purely real and describes the hidden order phase
of URu$_2$Si$_2$. This solution has zero magnetic moment, does not
break time reversal symmetry and has nonzero hexadecapole.
%with wave vector
%$Q=(0,0,1)$.
%
%It is characterized by a non zero staggered hexadecapole, a very high
%order multiple which is hard to detect experimentally.
%
The second DMFT solution, has a purely imaginary $\langle\psi\rangle$
and we associate this phase with the large moment antiferromagnet
phase, which is experimentally realized at pressures larger than
$0.7\,$GPa \cite{Vojta,Amitsuka,Motoyama,Flouquet_pressure}.  Hence,
the microscopic approach succeeds in unifying two very distinct broken
symmetry states in a single complex order parameter. The existence of
a solutions with either purely real or purely imaginary $\psi$, but
without mixtures, indicates a first order phase transition between
these two phases, as observed in the pressure experiments of
Ref.~\cite{Vojta,Amitsuka,Motoyama,Flouquet_pressure}.
%%% 301 words

It is useful to visualize the meaning of this order parameter in a
limiting case of a simple atomic wavefunction, where it takes the form
$|gs> = \cos(\theta)|\emptyset\rangle + \sin(\theta) e^{i{\varphi}}|1\rangle$
The average magnetic moment of the ground state is hence $\langle
gs|\textbf{J}|gs\rangle=4 \cos(\phi) \sin(2\theta) \sin({\varphi})
* (0,0,1)$.
If ${\varphi}$ is $\pi/2$, we have a phase with large magnetic moment,
and if ${\varphi}$ vanishes, the moment vanishes.

Our theory provides a natural explanation for a large number of
experiments, which are very puzzling, when examined from other
perspectives and suggests new experiments.

Firstly, it has been advocated phenomenologically that even though the
hidden order phase and the large moment phase have distinct order
parameters, the behavior of many observables across the transition is
remarkably similar. The term \textit{adiabatic continuity} has been
used to describe this situation \cite{Balicas,Vojta}, but it is not
justified on the theoretical grounds since the two phases are
separated by a first order phase transition.
The proposed order parameter $\psi_i$, which unifies the no-moment and
large moment phase, explains why even though the two phases are
separated by a first order phase transition, they are in many respects
very similar, for example in the critical temperature, and entropy
change across the transition.

Secondly, at the hidden order transition, a small gap of the order of
$10\,$meV opens in the $5f$ quasiparticle spectra, as seen in optical
conductivity \cite{optics}, specific heat \cite{cv_magnetic_field},
thermal conductivity \cite{Behnia}, and relaxation rate measurements
\cite{NMR,resistivity}.  On the other hand the dc resistivity
continues to decrease at low temperatures \cite{Hall}, since it is
dominated by the itinerant $spd$ carriers. This fact is hard to
understand in a simple itinerant density wave picture.
The $f$ states gap at low energies is also  seen in
neutron scattering experiments \cite{Broholm,Neutrons,Wiebe}.

Thirdly, unlike density functional theory calculations
\cite{Oppeneer}, our approach described a strongly correlated normal
state with large entropy, and can account for large specific heat
coefficient in the paramagnetic state of the material.

%% NEED CHANGE!!!
%Finally, the Gindzurg-Landau functional for the order parameter
%$\psi_i$ would contain coupling of its real part to the stress in the
%ab-plane $\sigma$, i.e.,
%%
%$\sum_i \Re\psi_i \sigma$, because the broken tetragonal symmetry
%mixes the two crystal field states.
%%
%The imaginary part of the order parameter would couple to the magnetic
%field as
%%
%$\sum_i \Im\psi_i h$.
%%
%This coupling at linear order  accounts for the sensitivity of the
%phase phase diagram to perturbation of stress \cite{stress} and
%magnetic field \cite{Balicas}.

It has recently been proposed that one can detect hexadecapole order
using resonant X-ray technique \cite{KURAMOTO}.
This would be a direct test of our proposed order parameter.
A high resolution angle resolved photoemission spectroscopy can detect
the small kink in the very low energy spectra ($<10\,$meV) of
Fig.~\ref{fig1}D, which is a unique signature of the arrested Kondo
effect.
%
% NEED CHANGE!!!
It would be interesting to control the crystal field splitting energy
$\Delta$. A slight decrease (increase) of $\Delta$, will increase
(decrease) the hidden order transition temperature.

Finally, our results set the stage for understanding the mysterious
superconducting transition that takes place at the much lower
temperatures ($T_c=0.8\,$K). Hidden order is fertile ground for
superconductivity, while the large moment antiferromagnetic phase
completely eliminates this instability. Our results suggest that
Cooper pairing can take place only when the electrons propagate in the
time reversal symmetric background, but unravelling the precise origin
of the superconductivity in this material will require further
sleuthing.
%%% 555

% I would not put acknowledgement yet, because we can not ask Davis
% and Allen to be referees. We will add acknowledgemnt if we get
% accepted.
%
%Acknowledgements: this work was supported by the NSF.  Useful
%discussions with J. Allen, S. Davis, J. Denlinger, P. Coleman.  P
%Chandra, are greatfully acknowledged.

\section{Acknowledgements}

We are grateful to J.~Allen and J.~Denlinger for fruitful discussion.
K.H was supported by Grant NSF NFS DMR-0746395 and Alfred P. Sloan
fellowship. G.K. was suported by NSF DMR-0906943.

%\section{Author contributions}
%K.H. and G.K. both developed the LDA+DMFT methodology
%and the physical interpretation of the results.
%
%\section{Additional information}
%Supplementary information accompanies this paper on
%www.nature.com/naturephysics.  Reprints and permissions information is
%available online at http://npg.nature.com/
%reprintsandpermissions. Correspondence and requests for materials
%should be addressed to K.H. or G.K.

\newpage
\begin{widetext}
  
\section{Online Material: Arrested  Kondo effect and hidden order in URu$_2$Si$_2$}

\subsection{The details of the method}

We carried out the band structure calculation for URu$_2$Si$_2$ using
recently developed method which combines the Density Functional Theory
(DFT) and Dynamical Mean Field Theory (DMFT). The combination DFT+DMFT
(for a review see Ref.~\cite{our-rmp}) contains some aspects of band
theory, adding a ``frequency-dependent local potential'' to the
Kohn-Sham Hamiltonian. It also contains some aspects of quantum
chemistry, carrying out an exact local configuration interaction
procedure by summing all local Feynman diagrams. The details of the
current implementation of the method are given in Ref.~\cite{newPaper}
(see in particular Eqs.(2) and (25) that give the site resolved
tunneling spectra).

The crystal structure of URu$_2$Si$_2$ is body-centred tetragonal with
point group I4/mmm (ThCr$_2$Si$_2$ type). The lattice parameters are
$a=4.1240\,\textrm{\AA}$, $c=9.5566\,\textrm{\AA}$ and the internal
parameter $z=0.3727$ \cite{struct}.

The on-site screened Coulomb repulsion in actinides ranges between
$4-5\,$eV. For uranium, which is on itinerant side of actindes, our
estimate for $U$ is $4.0\,$eV. The Slater integrals were computed by
atomic physics program of R.D. Cowan \cite{Cowan}, and scaled by 70\%
to account for screening in the solid. Their values are
$F_2=6.81\,$eV, $F_4=4.55\,$eV, $F_6= 3.36\,$eV.

For the double-counting, we used the standard, localized formula
$E_{DC}=U(n_{imp}-1/2)-J(n_{imp}/2-1/2)$ with
$J=F_2/11.92196532=0.57\,$eV, and $n_{imp}$ is the average number of
electrons in the U-$5f$ orbital, for URu$_2$Si$_2$ $n_{imp}\approx 2$.

We used two complementary impurity solvers, the One Crossing
Approximation (OCA) \cite{our-rmp,newPaper} and the Continuous Time
Quantum Monte Carlo (CTQMC) \cite{CTQMC}. The first is implemented
directly on the real axis and is of invaluable help to analytically
continue the exact Monte Carlo imaginary time data.
For the DFT part, we used the LMTO method of S. Savrasov
\cite{Savrasov} and we crosschecked results with LAPW method, as
implemented in Wien2K~\cite{Wien2K}.

The URu$_2$Si$_2$ compound is one of the rare cases, where the charge
self-consistency of DFT+DMFT is crucial. Namely, the DFT potential
needs to be computed on the self-consistent DFT+DMFT electronic
charge. The non-self consistent version of DFT+DMFT severely misplaces
the important low energy bands.

\subsection{The broken symmetry states}

At low temperature, we succeeded to stabilize two broken symmetry
states with the ordering wave wector $Q=(0,0,1)$: An antiferromagnetic
state with large magnetic moment, and a broken symmetry state with no
magnetic moment.

There are two independent ways to look at the nature of the broken
symmetry state: i) An atomic perspective, which was given in the main
text of the article. The broken symmetry solution is associated with
the local order parameter $\psi=\langle X_{\emptyset
  1}\rangle=\langle|\emptyset\rangle\langle 1|\rangle$, where
$X_{\emptyset 1}$ is the Hubbard operator, which mixes two crystal
field singlet states, the ground state and the first excited state of
the atom. The two atomic states are given by
\begin{eqnarray}
|\emptyset\rangle &=& \frac{i}{\sqrt{2}}(|4\rangle -|-4\rangle)\\
|1\rangle &=& \frac{\cos(\phi)}{\sqrt{2}}(|4\rangle +
|-4\rangle)-\sin(\phi)|0\rangle.
\end{eqnarray}
Here atomic states $|0\rangle$ and $|4\rangle$ correspond to the
$|J=4,J_z=0\rangle$ and $|J=4,J_z=4\rangle$, respectively.
ii) Alternatively, one can examine the structure of Green's function,
the central object of the DFT+DMFT method. To understand the symmetry
of the broken symmetry solution, one can examine the equal time
analog, which is the density matrix of the problem.

Below we will derive the connection between the density matrix of the
problem and the atomic eigenstates for the case of
URu$_2$Si$_2$. We first evaluate the projection of the
density matrix operator $f^\dagger_\alpha f_\beta$ to the atomic
eigenstates.
It is useful to view the broken symmetry from the point of view of the
Green's functions matrix and its equal time limit (the density matrix).
Exact diagonalization of the atomic problem leads to the following
relations
\begin{eqnarray}
\label{s0}
&f_{5/2}^\dagger f_{-3/2} &= -a |4\rangle\langle 0| + b |0\rangle\langle -4|+\cdots\\
\label{s1}
&f_{-3/2}^\dagger f_{5/2} &= -a |0\rangle\langle 4| + b |-4\rangle\langle 0|+\cdots \\
\label{s2}
&f_{3/2}^\dagger f_{-5/2} &= b |4\rangle\langle 0|-a |0\rangle\langle -4|+\cdots  \\
\label{s3}
&f_{-5/2}^\dagger f_{3/2} &= b |0\rangle\langle 4|-a |-4\rangle\langle 0|+\cdots   \\
\label{s4}
&f_{-5/2}^\dagger f_{-5/2} &= c_1 |-4\rangle\langle -4|+c_2 |0\rangle\langle 0|+\cdots   \\
\label{s5}
&f_{-3/2}^\dagger f_{-3/2} &= c_3 |-4\rangle\langle -4|+c_4 |0\rangle\langle 0|+\cdots   \\
\label{s6}
&f_{3/2}^\dagger f_{3/2} &= c_3 |4\rangle\langle 4|+c_4 |0\rangle\langle 0|+\cdots   \\
\label{s7}
&f_{5/2}^\dagger f_{5/2} &= c_1 |4\rangle\langle 4|+c_2 |0\rangle\langle 0|+\cdots   
\end{eqnarray}
with $a\approx  0.77$, $b\approx 0.25$, $c_1\approx 0.97$,
$c_2\approx 0.07$, $c_3\approx 0.98$, $c_4\approx 0.63$.
%Here atomic states $|0\rangle$ and $|4\rangle$ correspond to the
%$|J=4,J_z=0\rangle$ and $|J=4,J_z=4\rangle$, respectively.

The relation between the crystal field atomic eigenstates
($|\emptyset\rangle$, $|1\rangle$, $|2\rangle$) and the direct atomic
states $|J,J_z\rangle$ ($J_z=4,-4,0$) is
\begin{eqnarray}
\left[
\begin{array}{c}
  |\emptyset\rangle\\
  |1\rangle\\
  |2\rangle
\end{array}
\right]=
\begin{bmatrix}
  \frac{i}{\sqrt{2}} & \frac{-i}{\sqrt{2}} & 0\\
  \frac{\cos\Phi}{\sqrt{2}} & \frac{\cos\Phi}{\sqrt{2}} & -\sin\Phi \\
  \frac{\sin\Phi}{\sqrt{2}} & \frac{\sin\Phi}{\sqrt{2}} & \cos\Phi
\end{bmatrix}
\left[
\begin{array}{l}
  |4\rangle\\
  |-4\rangle\\
  |0\rangle
\end{array}
\right].
\end{eqnarray}
Inverting this matrix equation and inserting it into
Eqns.~(\ref{s0})-(\ref{s7}), we obtain the followin form of the density matrix
%hence the connection between the density matrix
%$\langle f^\dagger_\alpha f_\beta\rangle$
\begin{eqnarray}
&\langle f_{5/2}^\dagger f_{-3/2}  \rangle&= \frac{i\sin\Phi}{\sqrt{2}}(b \psi^*- a \psi)+\cdots=\frac{\sin\Phi}{\sqrt{2}}[i(b-a)\Re\psi+(b+a)\Im\psi]+\cdots\\
&\langle f_{-3/2}^\dagger f_{5/2}  \rangle&= \frac{i\sin\Phi}{\sqrt{2}}(a \psi^*- b \psi)+\cdots=\frac{\sin\Phi}{\sqrt{2}}[-i(b-a)\Re\psi+(b+a)\Im\psi]+\cdots\\
&\langle f_{3/2}^\dagger f_{-5/2}  \rangle&= \frac{-i\sin\Phi}{\sqrt{2}}(a \psi^*- b \psi)+\cdots=\frac{\sin\Phi}{\sqrt{2}}[i(b-a)\Re\psi-(b+a)\Im\psi]+\cdots\\
&\langle f_{-5/2}^\dagger f_{3/2}  \rangle&= \frac{-i\sin\Phi}{\sqrt{2}}(b \psi^*- a \psi)+\cdots=\frac{\sin\Phi}{\sqrt{2}}[-i(b-a)\Re\psi-(b+a)\Im\psi]+\cdots\\
&\langle f_{-5/2}^\dagger f_{-5/2} \rangle&= \frac{-i c_1 \cos\Phi}{2}(\psi^*- \psi)+\cdots=-c_1\cos\Phi\; \Im\psi+\cdots\\ 
&\langle f_{-3/2}^\dagger f_{-3/2} \rangle&= \frac{-i c_3 \cos\Phi}{2}(\psi^*- \psi)+\cdots=-c_3\cos\Phi\; \Im\psi+\cdots\\ 
&\langle f_{3/2}^\dagger f_{3/2}   \rangle&= \frac{i c_3 \cos\Phi}{2}(\psi^*- \psi)+\cdots=c_3\cos\Phi\; \Im\psi+\cdots\\ 
&\langle f_{5/2}^\dagger f_{5/2}   \rangle&= \frac{i c_1 \cos\Phi}{2}(\psi^*- \psi)+\cdots=c_1\cos\Phi\; \Im\psi+\cdots.
\end{eqnarray}
where $\psi$ is the order parameter
$\psi=\langle|\emptyset\rangle\langle 1|\rangle$.  Here the dots
($\cdots$) stand for the projection to the other atomic states, the
part of the projection which is nonzero in the paramagnetic state.

Setting $\Im\psi= 0$ and $Re\psi\ne 0$, we see that
$\delta \langle f^\dagger_{5/2}f_{-3/2}\rangle= -\delta \langle
f^\dagger_{-3/2}f_{5/2}\rangle$ and
$\delta \langle f^\dagger_{-5/2,-5/2}f_{-5/2,-5/2}\rangle=\delta \langle
f^\dagger_{5/2,5/2}f_{5/2,5/2}\rangle$. The symbol $\delta$ stands for
the change of the density matrix compared to its value in the
paramagnetic state.
The structure of the Green's function in the hidden-order phase is
therefore
\begin{eqnarray}
\Delta=\left(
\begin{array}{r|cccccc}
&-5/2 & -3/2 & -1/2 & 1/2 & 3/2 & 5/2 \\
\hline
-5/2& \Delta_A & 0 & 0 & 0 & \Delta_\varepsilon+\Delta_\alpha & 0\\
-3/2& 0 & \Delta_B & 0 & 0 & 0 & \Delta_\varepsilon+\Delta_\beta\\
-1/2& 0 & 0 & \Delta_C & 0 & 0 & 0\\
1/2 & 0 & 0 & 0 & \Delta_{C} & 0 & 0\\
3/2 & \Delta_\varepsilon-\Delta_\alpha & 0 & 0 & 0 & \Delta_{B} & 0\\
5/2 & 0 & \Delta_\varepsilon-\Delta_\beta & 0 & 0 & 0 & \Delta_{A}\\
\end{array}
\right)
\end{eqnarray}
Here $\Delta_{\varepsilon}$ is the off-diagonal term, present also in
the paramagnetic phase and comes due to the tetragonal crystal
environment.  Since the magnetic moment of this state vanishes
($\langle \textbf{J}\rangle=0$), we associate this moment-free phase
with the hidden order state.

The second solution has $\Im\psi\ne 0$ and $Re\psi=0$, hence
$\delta \langle f^\dagger_{5/2}f_{-3/2}\rangle= \delta \langle
f^\dagger_{-3/2}f_{5/2}\rangle$ and
$\delta \langle f^\dagger_{-5/2,-5/2}f_{-5/2,-5/2}\rangle\ne\delta \langle
f^\dagger_{5/2,5/2}f_{5/2,5/2}\rangle$.
%
%\subsubsection{Large moment phase}
The structure of the Green's function in the local moment
antiferromagnetic phase is therefore given by
\begin{eqnarray}
\Delta=\left(
\begin{array}{r|cccccc}
&-5/2 & -3/2 & -1/2 & 1/2 & 3/2 & 5/2 \\
\hline
-5/2& \Delta_A & 0 & 0 & 0 & \Delta_\varepsilon & 0\\
-3/2& 0 & \Delta_B & 0 & 0 & 0 & \Delta'_\varepsilon\\
-1/2& 0 & 0 & \Delta_C & 0 & 0 & 0\\
1/2 & 0 & 0 & 0 & \Delta'_{C} & 0 & 0\\
3/2 & \Delta_\varepsilon & 0 & 0 & 0 & \Delta'_{B} & 0\\
5/2 & 0 & \Delta'_\varepsilon & 0 & 0 & 0 & \Delta'_{A}\\
\end{array}
\right)
\end{eqnarray}
with $\Delta_{A}\ne\Delta'_{A}$, $\Delta_{B}\ne\Delta'_{B}$,
$\Delta_{C}\ne\Delta'_{C}$. This phase has nonzero magnetic moment
$\langle\emptyset|J|\emptyset\rangle=4$ and hence we associate it with
the large moment antiferromagnetic phase in URu$_2$Si$_2$.

%In this broken symmetry phase, the mixing between the two important
%singlet states
%\begin{eqnarray}
%|0\rangle &=& \frac{i}{\sqrt{2}}(|4\rangle -|-4\rangle)\\
%|1\rangle &=& \frac{\cos(\phi)}{\sqrt{2}}(|4\rangle +
%|-4\rangle)-\sin(\phi)|0\rangle
%\end{eqnarray}
%is possible and hence the pseudoparticle Green's function for the
%mixing is nonzero and has the form
%\begin{eqnarray}
%G_{01}=G_{10}\ne 0
%\end{eqnarray}
%
%
%\subsubsection{Moment free phase}
%
%In the hidden order phase, the Green's function and the hybridization
%function have the form
%\begin{eqnarray}
%\Delta=\left(
%\begin{array}{r|cccccc}
%&-5/2 & -3/2 & -1/2 & 1/2 & 3/2 & 5/2 \\
%\hline
%-5/2& \Delta_A & 0 & 0 & 0 & \Delta_\varepsilon+\Delta_\alpha & 0\\
%-3/2& 0 & \Delta_B & 0 & 0 & 0 & \Delta_\varepsilon+\Delta_\beta\\
%-1/2& 0 & 0 & \Delta_C & 0 & 0 & 0\\
%1/2 & 0 & 0 & 0 & \Delta_{C} & 0 & 0\\
%3/2 & \Delta_\varepsilon-\Delta_\alpha & 0 & 0 & 0 & \Delta_{B} & 0\\
%5/2 & 0 & \Delta_\varepsilon-\Delta_\beta & 0 & 0 & 0 & \Delta_{A}\\
%\end{array}
%\right)
%\end{eqnarray}

Another way to look at the broken symmetry state is by using the
perturbation theory around the atomic limit. In the expansion with
respect to the hybridization strength $\Delta$ (between the atom and the
conduction bands), the leading term of the Lutting-Ward functional
contains the following terms
\begin{equation}
\Phi = \sum_m c_m G_{mm} * \left[
  G_{01}*\left(\Delta_{-5/2,3/2}-\Delta_{3/2,-5/2}\right)+
  G_{10}*\left(\Delta_{-3/2,5/2}-\Delta_{5/2,-3/2}\right)
  \right]+\sum_m d_m G_{mm} * \left(G_{01}-G_{10}\right)*
  \left(\Delta_{j_z}-\Delta_{-j_z}\right)
\end{equation}
Here the pseudoparticle Green's function $G_{01}$ and $G_{10}$ are
connected to the Hubbard operators by
$G_{01}(\tau-\tau')=-\langle T_\tau X_{01}(\tau)X_{01}(\tau')\rangle$ and 
$G_{01}(\tau-\tau')=-\langle T_\tau X_{10}(\tau)X_{10}(\tau')\rangle$, respectively.

%The two broken symmetry solutions have nonzero off-diagonal Green's
%function $G_{01}$. The moment-free solution has  $G_{01}=G_{10}\ne 0$
%and the large moment solution has $G_{01}=-G_{10}\ne 0$.
%To check that this symmetry of the solution is possible, we evaluate
%the self-energies $\Sigma_{mn}=\delta\Phi/\delta G_{nm}$ and we get
%\begin{eqnarray}
%\Sigma_{01} = \sum_m c_m G_{mm}*\left(\Delta_{-5/2,3/2}-\Delta_{3/2,-5/2}\right)
%  +\sum_m d_m G_{mm} * \left(\Delta_{j_z}-\Delta_{-j_z}\right)\\
%\Sigma_{10} = \sum_m c_m G_{mm}*\left(\Delta_{-3/2,5/2}-\Delta_{5/2,-3/2}\right)
%  -\sum_m d_m G_{mm} * \left(\Delta_{j_z}-\Delta_{-j_z}\right).
%\end{eqnarray}

One possible self-consistent solution of this functional is $G_{01}=G_{10}\ne 0$,
$\left(\Delta_{-5/2,3/2}-\Delta_{3/2,-5/2}\right)=\left(\Delta_{-3/2,5/2}-\Delta_{5/2,-3/2}\right)\ne
0$ and $\Delta_{j_z}= \Delta_{-j_z}$. This solution corresponds to
moment-free phase.

The second possible solution is $G_{01}=-G_{10}\ne 0$,
$\Delta_{j_z}\ne \Delta_{-j_z}$ and
$\Delta_{-5/2,3/2}=\Delta_{3/2,-5/2}$,
$\Delta_{-3/2,5/2}=\Delta_{5/2,-3/2}$, which corresponds the
antiferromagnetic solution.

\end{widetext}

\end{document}